\documentclass{iaus}
\usepackage{graphicx}

\title[Window Functions in Planet Transit Searches] 
{Observational Window Functions in Planet Transit Searches}

\author[Kaspar von Braun \& David R. Ciardi]
{Kaspar von Braun$^1$
\and David R. Ciardi$^1$}

\affiliation{$^1$Michelson Science Center \\ California Institute of
Technology \\ 770 South Wilson Ave. \\ Pasadena, CA 91125, USA  
\\ email: {\tt kaspar@ipac.caltech.edu; ciardi@ipac.caltech.edu} \\}

\pubyear{2007}
\volume{249}  %% insert here IAU Symposium No.
\pagerange{1--6}
\jname{Exoplanets: Detection, Formation, and Dynamics}
\editors{Ji-Lin Zhou, ed.}
\begin{document}

\maketitle

% ---------------------------------------------------------------------

\begin{abstract}

Window functions describe, as a function of orbital period, the probability
that an existing planetary transit is detectable in one's data for a given
observing strategy. We show the dependence of this probability upon several
strategy and astrophysical parameters, such as length of observing run,
observing cadence, length of night, and transit duration.  The ability to
detect a transit is directly related to the intrinsic noise of the
observations. In our simulations of the window function, we explicitly address
non-correlated (white) noise and correlated (red) noise and discuss how these
two different noise components affect window functions in different manners.

\end{abstract}

% ---------------------------------------------------------------------

\firstsection % if your document starts with a section,
\section{Introduction}
\label{introduction}

The signal-to-noise ratio (SNR) of a planetary transit detection in a given
photometric light curve (here defined as simply time versus magnitude,
irrespective of presence of absence of variability) can, in the simplest case,
be approximated by:

\begin{equation}
SNR = \frac{depth}{\sigma}\sqrt{n}, 
\label{simple_equation}
\end{equation}

\noindent where {\it depth} is the transit depth in 
magnitudes, $\sigma$ represents the photometric measurement uncertainty in
magnitudes per data point (assumed here to be the same for all data points),
and {\it n} equals the number of data points observed during transit. The
fundamental assumption in this equation is the absence of any statistically
correlated (red) noise, i.e., the only source of noise is random (white)
noise.

White noise is defined as noise that is uncorrelated from data point to data
point; typical sources are photon noise and sky background noise. White noise
decreases with increasing brightness of the observed target.  Red noise is
defined as noise that is correlated from data point to data point; it is not
necessarily removed through standard differential or ensemble photometry
techniques. Typical sources of red noise may be weather, seeing changes, or
tracking errors. It does not change as a function of target magnitude.

As \cite{p06} and \cite{pzq06} initially pointed out, the assumption of the
presence of only white noise in equation \ref{simple_equation} is incorrect,
and one needs to account for the presence of red noise in the calculation of
the SNR and the corresponding expected yields for transit surveys, given
survey and astrophysical parameters. A more detailed description of the
transit detection SNR which includes both uncorrelated (white) and
correlated (red) noise components is given by \cite{pzq06}:

\begin{equation}
SNR = 
\frac{ depth }{\sqrt{\frac{1}{n^2} \sum_{i,j}C_{ij} }}
=
\frac{ depth }{\sqrt{\frac{\sigma^2}{n} + 
\frac{1}{n^2}\sum_{i\neq j}C_{ij} }},
\label{cov_equation}
\end{equation}
 
\noindent where $C_{ij}$ is the covariance matrix. In equation 
\ref{cov_equation}, the elements $C_{ij}$ represent the correlation 
coefficents between the $i$-th and $j$-th measurement obtained during
transit. All diagonal elements $C_{ii}=\sigma^2_i$ are not correlated with
other measurements and thus represent the uncorrelated or white noise
uncertainties in the $i$-th measurement. It is furthermore assumed that
$\sigma_i = \sigma$ for all values of $i$.

In order to make the above equation more practically calculable, \cite{pzq06}
assume that statistical correlation among data points from different transits
will be much weaker than among data points observed during the same
transit. They furthermore separate the total noise into a purely uncorrelated
(white) component $\sigma_w$ and a purely correlated (red) component
$\sigma_r$ and use these to derive an approximation of equation
\ref{cov_equation}:

\begin{equation}
SNR = \sqrt{
\frac{ \left ( depth \cdot n \right )^{2}}{\sum^{N_{tr}}_{k=1}\left[ n^2_k \left ( \frac{\sigma^2_w}{n_k} +
\sigma^2_r \right ) \right ] } },
\label{approximation_equation}
\end{equation}

\noindent where $n$ is the total number of data points observed during all
transits, $N_{tr}$ is the total number of transits observed, $n_{k}$ is the
number of data points observed during the $k$-th transit, and $\sigma_w$ and
$\sigma_r$ are the white and red noise components, respectively.

By means of equation \ref{approximation_equation}, the SNR can be regarded as
a function of transit survey strategy and astrophysical parameters (see \S
\ref{algorithm}). If it exceeds a certain threshold value, then an existing
transiting planet is detectable in the data. The window function indicates the
probability, as a function of orbital period, that the SNR ratio exceeds this
threshold and thus produces a detectable planet transit.

We briefly outline our algorithm and methods in \S 2, show respective
influences of varying white and red noise components in \S 3, and examine the
effects of various survey strategy and astrophysical parameters in \S4. We
summarize and conclude in \S 5. 

% ---------------------------------------------------------------------

\section{Algorithm and Input Parameters}
\label{algorithm}

The algorithm used for the calculation of the window functions uses input on
observing cadence as well as the number and typical length of night to
generate an observing time line. From the user-provided stellar and planetary
radii, it calculates transit depth and duration according to the equations in
\cite{sm03}, thereby assuming a central transit.  For each orbital period, a
family of light curves is generated for a range of starting phase angles.
Finally, the magnitude of the white and red noise components ($\sigma_{w}$ and
$\sigma_{r}$, as defined in \S \ref{introduction} and \cite[Pont et
al. 2006]{pzq06}) are specified in the input.

In the simulations, the number of data points per transit ($n_k$), number of
transits ($N_{tr}$) and total number of data points within all transits ($n$)
are tracked. For every light curve, the SNR (equation
\ref{approximation_equation}) is calculated. If, for a given phase angle, the
SNR exceeds SNR$_{threshold}$, a transit is considered ``detected''. The
probability of detection (P$_{Detection}$) for a given orbital period is
simply the ratio of phase angles for which a transit was detected to the total
number of phase angles.

We attempted to choose the input values such that they resemble the ones found
in transit surveys described in the literature, though our choice is probably
slightly tilted toward surveys that do not have the luxury of having dedicated
survey facilities. Thus, typical observational parameter values are: tens of
minutes for the observing cadence, tens of nights for observing run length,
and few to ten hours for the typical time of observation spent during one
night on the monitored target. Astrophysical parameter values are assumed to
be around 1.0 and 0.1 solar radii for the parent star and orbiting planet,
respectively, resulting in a transit depth of 0.01 mag (transit duration
depends on period, but typical duty cycles are in the 1\% to few \% range),
and a few millimagnitudes (mmag) for $\sigma_w$ and $\sigma_r$. The threshold
SNR is set to 10.

We note that, in contrast to other window function calculations, we only use
the SNR criterion to quantify detections, and do not require that, e.g., a
full transit be contained in the data or that data from at least two or more
transits be sampled.

% ---------------------------------------------------------------------

\section{Red Noise versus White Noise}
\label{noise_comparison}

\begin{figure}
% \vspace*{-2.0 cm}
\begin{center}
 \includegraphics[angle=90,scale=0.4,keepaspectratio=true ]{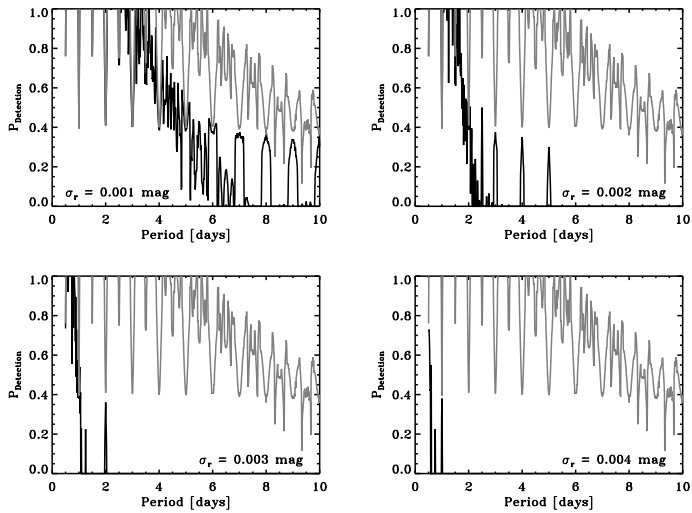} 
% \vspace*{-1.0 cm}
 \caption{$\sigma_w$ constant (5 mmag), $\sigma_r$ variable. The grey line
 (identical in all four panels) corresponds to the detection probability in
 the absence of red noise ($\sigma_r = 0$). The black line shows the same
 probability for varying levels of $\sigma_r$, indicated in each of the four
 panels. Other input parameters are SNR$_{threshold}$ = 10, stellar/planetary
 radius = 1.0/0.1 R$_{sun}$, 30 consecutive observing nights, 8 hours of
 observing per night, and an observing cadence of 10 minutes.}  \label{fig1}
\end{center}
\end{figure}

\begin{figure}
% \vspace*{-2.0 cm}
\begin{center}
 \includegraphics[angle=90,scale=0.4,keepaspectratio=true]{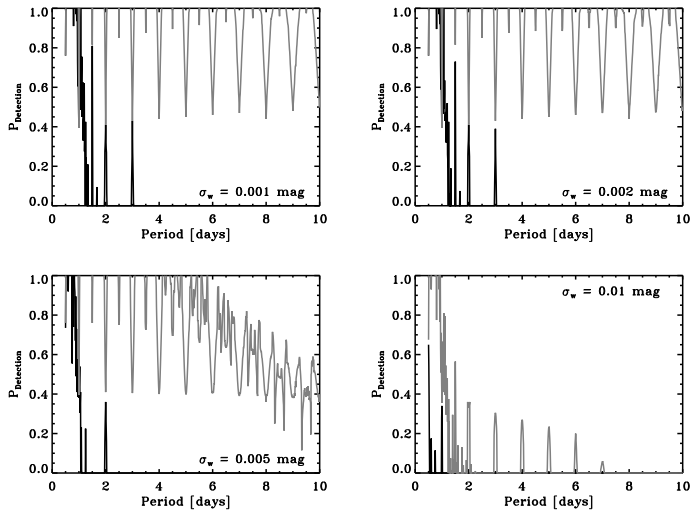} 
% \vspace*{-1.0 cm}
 \caption{$\sigma_r$ constant (3 mmag), $\sigma_w$ variable (shown in each of
 the four panels). As in Fig. \ref{fig1}, the grey line corresponds to the
 detection probability in the absence of red noise, for the purpose of
 comparison with the actual window function when red noise is present (black
 line). Other input parameters are the same as in Fig \ref{fig1}. Clearly, the
 $\sigma_r$ dominates, even in the case of exceptionally low white noise.}
 \label{fig2}
\end{center}
\end{figure}

Due to their low signal depths ($<1\%$), transits can typically only be
detected for the bright stars in one's sample.  For the brightest stars within
a survey, photon noise generally dominates the white noise. However,
\cite{pzq06} show that red noise (independent of target brightness) is
particularly important in this regime.  A more quantitative justification of
this statement can be found in \cite{ap07}, based on arguments initially
presented in \cite{pg05}. In Figs. \ref{fig1} and \ref{fig2}, we show the
effects of varying amounts of white and red noise on the window function.
Input parameters (as described in \S \ref{algorithm}) are given in the
captions of all figures.  The grey line in the figure panels always indicates
the window function in the absence of any correlated noise ($\sigma_r =
0$). In comparison, the black line indicates the window function resulting
from presence of both white and red noise ($\sigma_r\neq0$).

Fig. \ref{fig1} shows the window function for a constant white noise term
$\sigma_w$ = 5 mmag in combination with a variable red noise
term. \cite{pzq06} quote the OGLE survey's brightest stars' $\sigma_w$ to be
around 5 mmag and $\sigma_r$ to be around 3--4 mmag. \cite{k07} quote the WASP
survey's $\sigma_r$ to be 2--3 mmag, which reduces to 1.5 mmag after
detrending using the SYSREM algorithm described in \cite{tmz05}. It is
impressive to see the decrease in the detection efficiency even for small
changes in $\sigma_r$ in Fig. \ref{fig1}, illustrating the importance of the
use of detrending algorithms for transit survey data, such as SYSREM
(\cite[Tamuz et al. 2005]{tmz05}) or TFA (\cite[Kov{\'a}cs et
al. 2005]{kbn05}).

Fig. \ref{fig2} shows the window function for a constant red noise term
$\sigma_r$ = 3 mmag in combination with a variable $\sigma_w$. For comparison
purposes, the case for $\sigma_r = 0$ is also shown (grey line) in each panel.

These Figures show that even small amounts of red noise will be the dominant
factor in any calculation concerning transit detectability or survey yields,
at least before any detrending is applied.

% ---------------------------------------------------------------------

\section{The Influence of Observing Strategy and Astrophysical 
Parameters on the Window Function}
\label{strategy}

The careful consideration and simulation of individual aspect of transit
survey strategy can significantly improve the detection efficiency, as
outlined in, e.g., \cite{msy03}, \cite{pg05}, and \cite{bls05}. Furthermore,
individual astrophysical parameters can have an influence on the window
function and associated detection probability. In this Section, we show the
effects of a number of different parameters, both in the presense (black line)
and absence (grey line) of red noise.

% ---------------------------------------------------------------------

\subsection{Observing Run Length}
\label{run_length}

Transit surveys that are not able to use their own dedicated equipment will
often be limited by the number of observing nights allocated to their
programs. Understanding whether or not one will reach a certain probability of
transit detectability in one's data after a given number of nights is
important. Furthermore, one may consider doubling the number of monitored
stars by dividing a long observing run into two parts and changing targets
after a certain number of nights. An estimate of how much the change in number
of observing nights influences the projected survey yield is crucial to the
design of a successful transit survey.

\begin{figure}
% \vspace*{-2.0 cm}
\begin{center}
 \includegraphics[angle=90,scale=0.4,keepaspectratio=true]{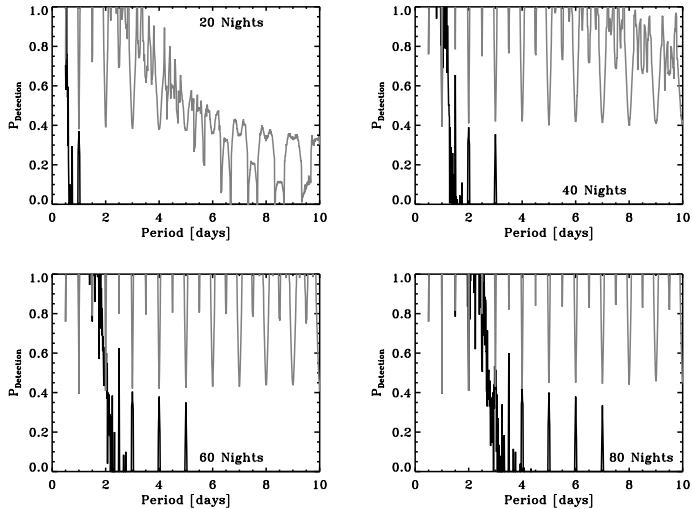}
% \vspace*{-1.0 cm}
 \caption{Length of observing run measured in number of 8-hour-long
 consecutive nights (indicated in each panel), both in the absence (grey line)
 and presence (black line) of red noise ($\sigma_r$ = 3 mmag). Other input
 parameters are a 10-min observing cadence, 8-hour-long nights, $\sigma_w$ = 5
 mmag, SNR$_{threshold}$ = 10, and stellar/planetary radius = 1.0/0.1
 R$_{sun}$. Changing the observing run length has a signficant effect on the
 transit detection efficiency.} \label{fig3}
\end{center}
\end{figure}

Fig. \ref{fig3} shows the detection probability for a range in observing run
lengths, measured in number of consecutive nights of 8 hours in length with an
observing cadence of 10 min. This confirms earlier results in the literature,
e.g., \cite{bls05} (based on zero red noise assumptions) that target field
switching during observing runs can be detrimental for projected survey yields
despite the increase in monitored stars.

% ---------------------------------------------------------------------

\subsection{Observing Cadence}
\label{cadence}

For a given telescope/detector combination, the observing cadence would change
as a function of target brightness (to reduce white noise and avoid
saturation). One could alternatively, for a given exposure time, conceive of a
survey strategy where one would move back and forth between pointings to
increase the number of monitored stars (as an alternative to \S
\ref{run_length} for, e.g., two separate fields that are near each other on
the sky).

\begin{figure}
% \vspace*{-2.0 cm}
\begin{center}
 \includegraphics[angle=90,scale=0.4,keepaspectratio=true]{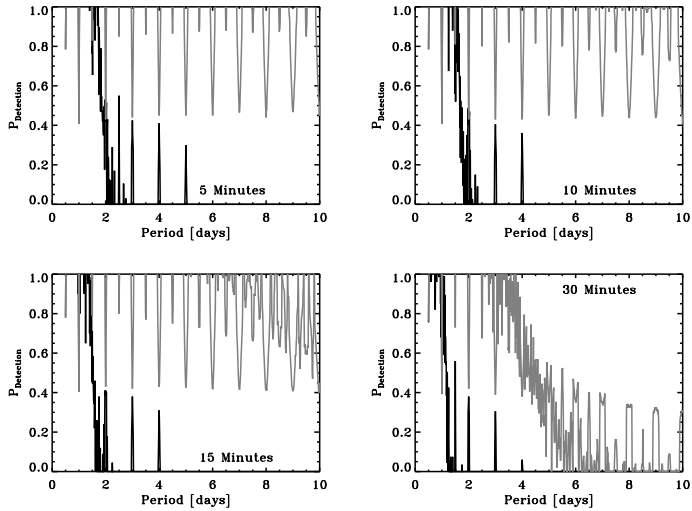} 
% \includegraphics[angle=-90, width=3.5in]{figures/cadence.eps} 
% \vspace*{-1.0 cm}
 \caption{Observing cadence with (black line) and without (grey line) red
 noise ($\sigma_r$ = 3 mmag). The effect of changing observing cadence is more
 pronounced in the white dominated regime. Other input parameters are 50
 consecutive 8-hour-long nights of observing, $\sigma_w$ = 5 mmag,
 SNR$_{threshold}$ = 10, and stellar/planetary radius = 1.0/0.1 R$_{sun}$.}
 \label{fig4}
\end{center}
\end{figure}

Fig. \ref{fig4} shows that the change in cadence greatly affects white noise
dominated regimes (grey line for $\sigma_r = 0$), but shows a relatively weak
influence when red noise is present. It should be pointed out, however, that
frequently changing between targets during the night will most likely
introduce a red noise component by itself.

% ---------------------------------------------------------------------

\subsection{Length of Night}
\label{night_length}

\begin{figure}
% \vspace*{-2.0 cm}
\begin{center}
 \includegraphics[angle=90,scale=0.4,keepaspectratio=true]{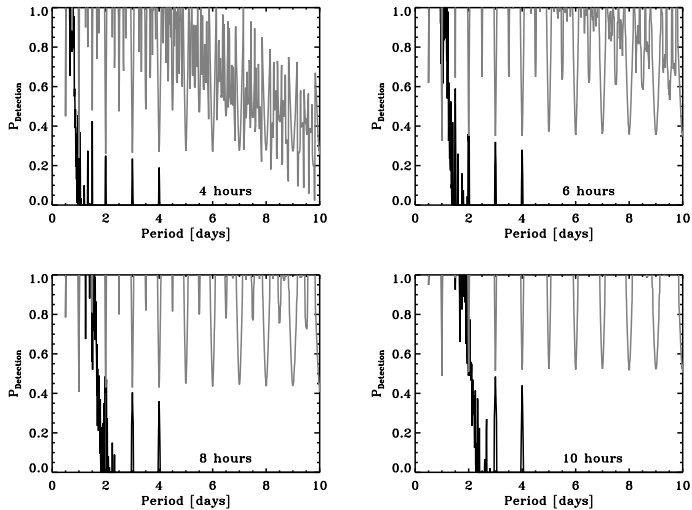} 
% \includegraphics[angle=-90, width=3.5in]{figures/night_length.eps} 
% \vspace*{-1.0 cm}
 \caption{Length of night in continuous hours of observing (indicated in all
 four panels) in the absence (grey line) and presence (black line) of red
 noise ($\sigma_r$ = 3 mmag).  Other input parameters are 50 consecutive
 nights of observing, a 10-min observing cadence, $\sigma_w$ = 5 mmag,
 SNR$_{threshold}$ = 10, and stellar/planetary radius = 1.0/0.1 R$_{sun}$.
 Changing the number of observing hours per night greatly influences transit
 detection probability.}  \label{fig5}
\end{center}
\end{figure}

The number of hours that a target field can be observed during the course of a
night depends on the combination of its coordinates, the location of the
observatory, and the time of year. Weather can also be a (more random)
factor. Finally, one may choose, again, to increase the number of monitored
stars by splitting the observing up into multiple target fields during the
night (as an alternative to the strategies described in \S \ref{run_length}
and \S \ref{cadence}). The effect of changing the length of night upon
detection efficiency is illustrated in Fig. \ref{fig5}, showing that, e.g.,
splitting the night in half to double the number of monitored stars noticeably
decreases the detection probability, especially for orbital periods longer
than 2 days, again in agreement the (white noise only) findings in
\cite{bls05}.

% ---------------------------------------------------------------------

\subsection{Transit Duration}
\label{transit_duration}

\begin{figure}
% \vspace*{-2.0 cm}
\begin{center}
 \includegraphics[angle=90,scale=0.4,keepaspectratio=true]{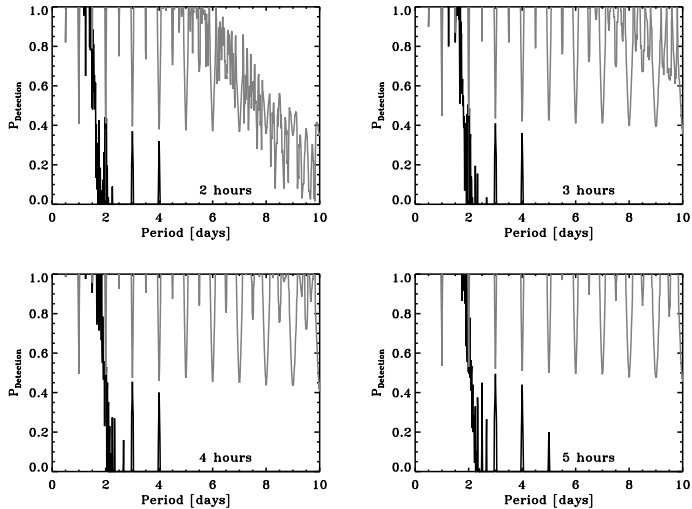} 
% \includegraphics[angle=-90, width=3.5in]{figures/transit_duration.eps} 
% \vspace*{-1.0 cm}
 \caption{Transit duration in hours (shown in each of the four panels), in the
 absence (grey line) and presence (black line) of red noise ($\sigma_r$ = 3
 mmag).  Other input parameters are 50 consecutive nights of observing, a
 10-min observing cadence, $\sigma_w$ = 5 mmag, SNR$_{threshold}$ = 10, and a
 transit depth of 0.01 mag. The extent to which the detection probability is
 affected by transit duration is higher for white noise dominated data.}
 \label{fig6}
\end{center}
\end{figure}

The duration of a transit is dependent upon the combination of stellar and
planetary radii, the orbital period, and the inclination angle of the system,
as described in, e.g., \cite{ma02} and \cite{sm03}. To simulate the effect of
a non-central transit, Fig. \ref{fig6} shows how the detection probability
changes as the duration of transit varies between 2 and 5 hours, typical for
planets with periods of around a few days. The impact of shorter transit
durations is noticeably higher in the white noise dominated regime (grey
line). This appears to indicate that, as long as the system actually transits,
the dependence of the detection efficiency upon inclination angle is
relatively weak.

% ---------------------------------------------------------------------

\section{Summary and Conclusions}

In this presentation, we illustrate the influence of several parameters on the
probability that an existing planetary transit is detectable in a data set
with given noise properties. Red noise dominates in the regime in which
planets are typically found (the brightest stars in one's sample). In order to
beat down red noise effects and improve detection efficiency, detrending is a
vital instrument.

Transit survey strategy can be employed to maximize the projected yield of a
given survey. We examine how much the detection efficiency for different
orbital periods would suffer when changing one's observing strategy, e.g., to
increase the number of monitored stars. We show examples involving the number
of consecutive observing nights, typical night length, as well as the
observing cadence, and we find that sacrificing full nights or parts of nights
can significantly lower transit detection probability. Finally, we show the
effects of non-central transits and the associated change of transit duration
upon detection efficiency. Our results indicate that, for $\sigma_r =$ 3 mmag
(typically found in some of the major transit surveys before detrending is
applied), small changes in inclination angle and associated transit duration
do not greatly affect detection efficiency.

In all simulations presented here, the presence of red noise (for $\sigma_r$ =
3 mmag) dominates all other effects. However, the variation of some of the
parameters examined in this work (in particular the observing cadence and the
transit duration) appears to have a bigger effect on white noise dominated
(e.g., detrended) data.

\end{document}